\documentclass[aip,pop,reprint,graphicx,superscriptaddress]{revtex4-1}
\usepackage{graphicx}
\usepackage{CJK}
\usepackage[dvipdfm,colorlinks,linkcolor=blue,citecolor=blue,urlcolor=blue]{hyperref}

\draft 

\begin{document}
\begin{CJK}{UTF8}{gbsn}


\title{High field terahertz emission from relativistic laser-driven plasma wakefields} 




\author{Zi-Yu Chen}
\email[Electronic mail: ]{Ziyu.Chen@uni-duesseldorf.de}
\affiliation{Institut f\"ur Theoretische Physik I, Heinrich-Heine-Universit\"at D\"usseldorf, D\"usseldorf 40225, Germany}
\affiliation{LSD, Institute of Fluid Physics, China Academy of Engineering Physics,
Mianyang 621999, China}

\author{Alexander Pukhov}
\affiliation{Institut f\"ur Theoretische Physik I, Heinrich-Heine-Universit\"at D\"usseldorf, D\"usseldorf 40225, Germany}
%


\date{\today}

\begin{abstract}
We propose a method to generate high field terahertz (THz) radiation
with peak strength of GV/cm level in the THz frequency gap range 1-10
THz using a relativistic laser interaction with a gaseous plasma target.
Due to the effect of local pump depletion, an initially Gaussian laser
pulse undergoes leading edge erosion and eventually evolves to a state
with leading edge being step function. Interacting with such a pulse,
electrons gain transverse residual momentum and excite net transverse
currents modulated by the relativistic plasma frequency. These currents
give rise to the low frequency THz emission. We demonstrate this process
with one and two dimensional particle-in-cell simulations. 
\end{abstract}

\pacs{}

\maketitle 


\section{Introduction}
High field terahertz (THz) pulses with peak electric field strength
$>1$ MV/cm in the previously hardly accessible THz frequency gap
range 1-10 THz have gained wide scientific interest, since they offer attractive opportunities
for studies in condensed matter physics, biology, chemistry and photonics.
As a unique and versatile tool, high field THz sources not only allow
probe and observe the state of matter with high sensitivity, but also
open a door to manipulate and control the electronic, ionic and spin
degrees of freedom of matter both resonantly and nonresonantly. \cite{Kampfrath2013}

Currently, high field THz pulses can be generated from large-scale
accelerators using ultrashort electron bunches. For example, peak
field of 44 MV/cm has been obtained via coherent transition radiation
in the Linac Coherent Light Source (LCLS).\cite{Wu2013} In the laser-based
schemes, THz pulses with field strength $>8$ MV/cm at 1 KHz repetition
rate have been generated via two-color laser filamentation.\cite{Oh2014}
Single-cycle THz pulse with field strength up to 36 MV/cm can be generated
by optical rectification of a midinfrared laser in a large-size nonlinear
organic crystals assembly.\cite{Vicario2014} In addition, other methods
that have produced high field THz pulses include difference frequency
mixing process of two near-infrared lasers in second-order nonlinear
crystals, but only with components $>20$ THz,\cite{Sell2008} and
optical rectification in lithium niobate (LiNbO$_{3}$) crystals with
tilted laser front, but mostly limited to $<1.5$ THz.\cite{Stepanov2014}
Besides, high field THz source can be generated from relativistic
laser irradiated plasmas via various mechanisms.\cite{Sheng2005,Li2012,Gopal2013,Chen2013a}

In our previous work, we have shown numerically that THz pulses with
field strength $>$ GV/cm can be generated in plasma wakefieds by
use of temporally tailored laser pulses.\cite{Chen2013b} The mechanism
is related to the excitation of transverse plasma current via electron
residual momentum left by the temporally tailored laser pulse. For
an electron initially at rest, the transverse momentum $p_{z}$ normalized
by $m_{e}c$ after the interaction with a $z$-polarized laser pulse
can be obtained as an integral of the transverse field $E_{z}$:\cite{Galow2011}
\begin{equation}
p_{z}=a_{z}=e/(m_{e}c)\int_{-\infty}^{+\infty}E_{z}(\eta)d\eta,\label{pzeq}
\end{equation}
where $a_{z}$ is the normalized vector potential, $\eta=t-x/c$,
$e$ is the elementary charge, $m_{e}$ is the electron mass, and
$c$ is the speed of light in vacuum. After the interaction with a
normal Gaussian pulse, the above integral is zero, and hence electrons
gain no net energy. However, if the laser pulse has such a steepened
edge that it becomes a step function, a large transverse residual
momentum $p_{z}^{R}$ as well as vector potential $a_{z}^{R}$ will
be generated. Considering the one-dimensional (1D) wave equation\cite{Lichters1996}
\begin{equation}
\Big(\partial_{x}^{2}-\frac{1}{c^{2}}\partial_{t}^{2}\Big)a_{z}^{T}=\Big(\frac{\omega_{p0}}{c}\Big)^{2}s_{z}(x,t)
\end{equation}
with the source term 
\begin{equation}
s_{z}(x,t)=\frac{n_{e}(x,t)a_{z}^{R}(x,t)}{\gamma(x,t)},
\end{equation}
where $a_{z}^{T}$ is the vector potential of the THz wave, $\omega_{p0}=(4\pi n_{0}e^{2}/m_{e})^{1/2}$
is the background plasma frequency, $n_{e}(x,t)$ is the plasma density
normalized by the initial density $n_{0}$, and $\gamma$ is the relativistic
Lorentz factor. The normalized THz electric field by $m_{e}\omega_{0}c/e$
can then be obtained as 
\begin{equation}\label{ezint}
e_{z}^{T}(x,t)=-\frac{1}{\omega_{0}}\partial_{t}a_{z}^{T}=\frac{\omega_{p0}}{2\omega_{0}}\int_{-\infty}^{+\infty}\frac{dx'}{l_{s}}s_{z}(x',t-|x-x'|/c),
\end{equation}
where $l_{s}=c/\omega_{p0}$ and $\omega_{0}$ is the laser angular
frequency.

In this work, we demonstrate that THz emission from laser plasma wakefields can be self-consistently realized with a normal Gaussian laser pulse, unlike previously the laser pulse with a step-function edge is introduced artificially.\cite{Chen2013b} 
The way to achieve the required pulse shape is to utilize the effect of localized
etching. When a high intensity laser pulse propagates in underdense
plasmas, the leading edge of the laser pulse can get eroded and depleted
via the local pump depletion mechanism,\cite{Decker1996} and thus
eventually evolves to a state with a step-function front. The net transverse current exited by such a pulse in the plasma wakefields, modulated by the relativistic plasma frequency, should be capable of generating low frequency THz pulses. 

\section{Simulation results and discussions}
To illustrate the THz generation process, 1D particle-in-cell (PIC) simulations have been
carried out using the Virtual Laser Plasma Laboratory (VLPL) code.\cite{Pukhov1999}
A circularly polarized laser with a Gaussian temporal profile $a(t)=a_{0}\exp(-t^{2}/\tau_{0}^{2})$
is normally incident along the $x$-axis, where $a_{0}$ and $\tau_{0}$
are respectively the laser normalized peak amplitude and the pulse
duration. We note that linearly polarized laser also works. Here, we take $a_{0}=eE_{0}/m_{e}c\omega_{0}=30$ with $E_{0}$
the laser electric field amplitude. The pulse duration is about 11
fs full-width at half-maximum (FWHM). The laser wavelength is taken
to be $\lambda_{0}=0.8\mu$m. A fully ionized plasma is initially
located between $50\lambda_{0}$ and $350\lambda_{0}$. For the laser intensities and plasma densities we considered here, the usual gas jets routinely used in experiments can be employed, such as hydrogen, helium and nitrogen gases.\cite{Huang2014,Yan2014} For simplicity,
a homogeneous plasma slab with a density of $n_{0}=0.01n_{c}$ is
used, where $n_{c}=m_{e}\omega_{0}^{2}/4\pi e^{2}\sim10^{21}$ W/cm$^{2}$
is the critical plasma density. We also checked inhomogeneous density
profiles with density gradients and find this mechanism also works.
The simulation box length is $550\lambda_{0}$. The grid size is $\lambda_{0}/100$
with each cell filled with 10 macroparticles.

\begin{figure}
\includegraphics[width=0.5\textwidth
]{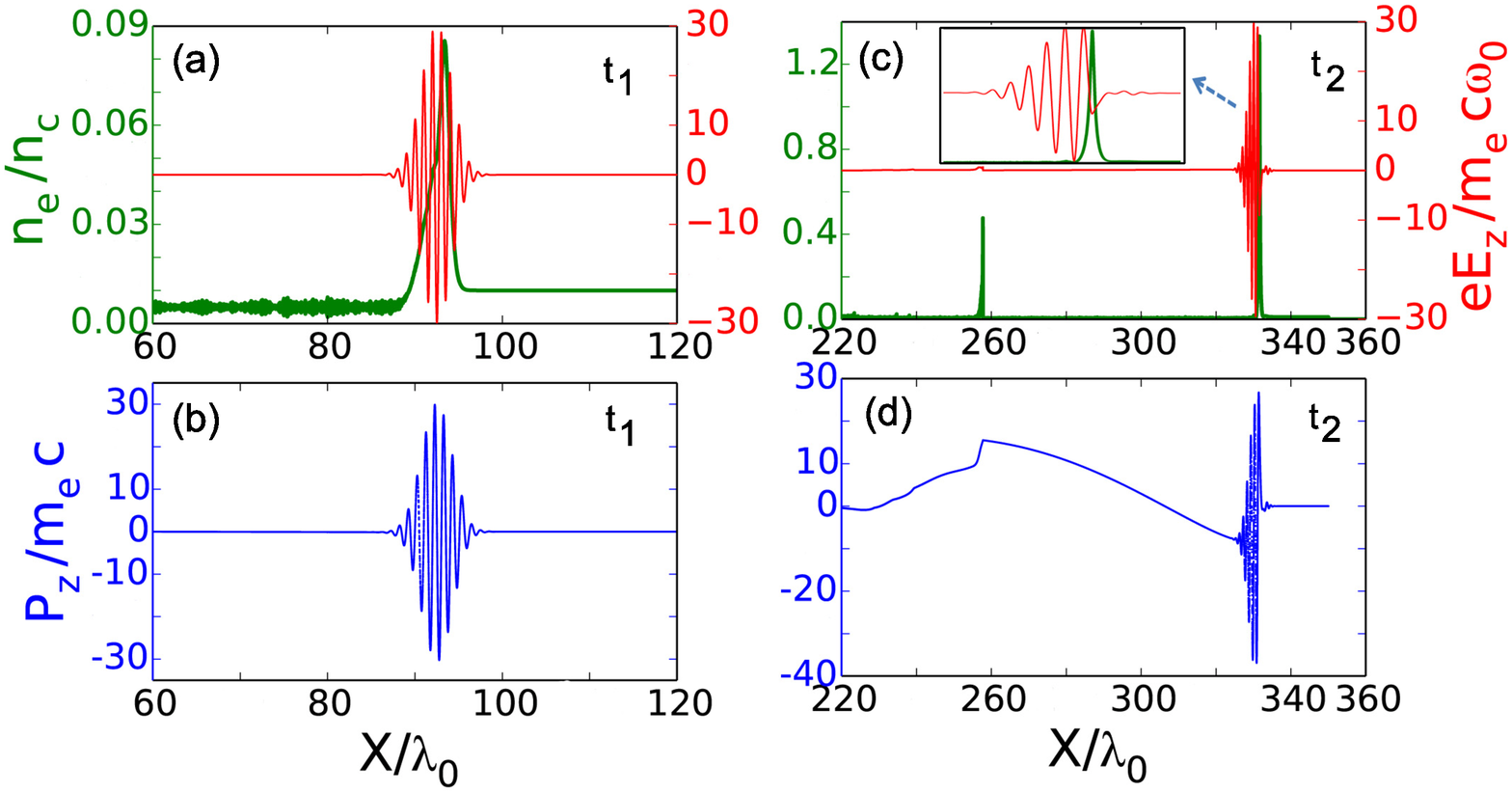}
\caption{\label{pz} (Color online) Spatial profiles of the electron density $n_e$ (green) and transverse field $E_z$ (red) at (a) $t_1=80 T_0$ and (c) $t_2=318T_0$. (b) and (d) are the corresponding spatial distribution of the transverse momentum $P_z$ for (a) and (c), respectively. The inset of (c) shows a close-up for the laser pulse shape.}
\end{figure}

Figure~\ref{pz} shows the snapshots of distribution of the laser
pulse $E_{z}$ and the electron density $n_{e}$ as well as the transverse
momentum of electron $P_{z}$ at two different times: $t_{1}=80T_{0}$
and $t_{2}=318T_{0}$, where $T_{0}$ is the laser period. In Fig.~\ref{pz}(a),
an electron density peak is formed at the front of the laser pulse
by snow plowing the electrons forward due to the laser ponderomotive
force. The peak electron density is larger than the initial background
density, but still much smaller than the critical density. The laser
pulse almost holds its initial symmetric temporal shape. The corresponding
transverse momentum $P_{z}$ in Fig.~\ref{pz}(b) only exits inside
the laser pulse. Behind the laser pulse, one finds $P_{z}\approx0$.
From Fig.~\ref{pz}(c), however, one can see that at a later time,
the density peak at the front of the pulse becomes narrow and sharp.
The amplitude of the density spike is above the critical density.
This results in an efficient erosion of the leading edge of the pulse
due to localized energy depletion mechanism within the narrow density
spike region.\cite{Decker1996} As the laser pulse propagates, the
density spike is continuously pushed by the laser ponderomotice force
in front of the pulse. Consequently, the laser front coinsides with
the density spike where it depletes its pump energy. Due to the etching
of the front, the laser pulse evolves to a state with an ultra-steep
front which is approximately a step function (see a close-up for the
laser pulse shape in the inset of Fig.~\ref{pz}(c)). As the laser
front continuously etches backwards, the sharp shape of the leading
edge is kept. Although there are some other effects like group velocity
steepening that can lead to the formation of steep pulse front,\cite{Vieira2010}
these processes are not so efficient. With this pulse shape of step-function
front, the integral in Eq.~(\ref{pzeq}) is no longer zero. As a
result, the electrons acquire a net transverse momentum after the
interaction with the laser field. This residual momentum left behind
the laser pulse can be clearly seen in Fig.~\ref{pz}(d). Due to
the residual momentum, the plasma electrons keep on a free oscillation
with the background plasma frequency $\omega_{p}$. Then
a low frequency transverse net current is built up in the plasma wakefields
and emits both forward and backward low frequency electromagnetic
radiation through the plasma slab.\cite{Wu2008} We noth that the second density peak at around 260 $\lambda_0$ behind the laser pulse in Fig.~\ref{pz}(c) is a trapped electron sheet injected into the wakefield. This electron sheet may generate an ultrafast XUV pulse in the presence of the residual transverse momentum.\cite{Liu2012}

\begin{figure}
\includegraphics[width=0.5\textwidth
]{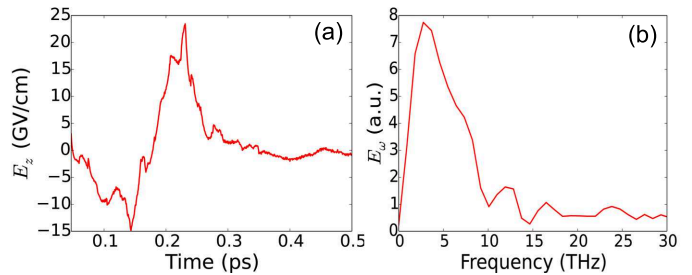}
\caption{\label{tf} (Color online) (a) Temporal waveform and (b) the corresponding Fourier spectrum of the forward THz emission field $E_z$ observed in front of the right plasma-vacuum boundary.}
\end{figure}

The temporal waveform and the corresponding Fourier spectrum of the
low frequency emission observed in the forward direction are shown
in Figs.~\ref{tf}(a) and (b), respectively. A single-cycle THz pulse
is obtained, which lasts for a few hundred femtoseconds. The peak
field strength of the THz pulse reaches about 24 GV/cm. The spectral
shape shows that most of the pulse energy concentrates in the desired
THz frequency gap range of 1-10 THz, with a peak frequency of about
3 THz.

\begin{figure}
\includegraphics[width=0.5\textwidth
]{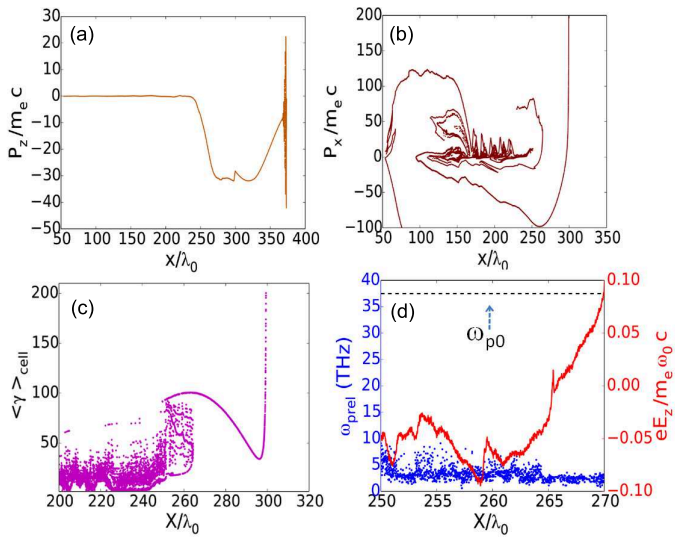}
\caption{\label{omega} (Color online) (a) $x-p_z$ and (b) $x-p_x$ phase space distributions of the electrons. (c) Spatial distribution of the relativistic $\gamma$-factor. (d) Spatial profiles of the relativistic plasma frequency $\omega_\mathrm{prel}$ (blue) and the transverse electric field $E_z$ (red) in the skin layer. The black dashed line marks the initial plasma frequency $\omega_{p0}$. The time is $t=360T_0$.}
\end{figure}

The initial plasma frequency is $\omega_{p0}\approx$37.5 THz, which
is greater than the central frequency of the observed THz emission.
This can be attributed to the change of the plasma frequency due to
relativistic effects. As a result of the interaction with the relativistic
laser, electrons reach near light speed and gain a large Lorentz factor
($\gamma=1/\sqrt{1-(v/c)^{2}}$). Therefore, the mass of electron
is increased and the effective plasma frequency is reduced. The relativistically
corrected expression for the plasma frequency is $\omega_{\mathrm{prel}}=(4\pi e^{2}n_{e}(x)/\gamma m_{e})^{1/2}$.
We plot the electrons' $\gamma$-factor along with the phase space
distribution at time $t=360T_{0}$ in Figs.~\ref{omega} (a)-(c).
The $\gamma$-factor is obtained by averaging the values in each cell,
i.e., $\gamma=<\gamma>_{\mathrm{cell}}$. In the skin layer near the
plasma rear surface, large residual transverse momentum can be seen,
which can excite net transverse currents radiating electromagnetic
pulses (see Fig.~\ref{omega}(a)). In this plasma region, the Lorentz
factor $\gamma$ of the bulk electrons reaches up to 100 (see Fig.~\ref{omega}(c)).
In addition to the bulk electrons, there are some high-energy electrons
with $\gamma>100$, which can also be seen in Fig.~\ref{omega}(b).
These electrons are refluxing backwards in the electrostatic field
due to charge separation with the ions. Unlike the bulk electrons,
they contribute little to the plasma frequency. Figure~\ref{omega}(d)
shows the profiles of the relativistic plasma frequency $\omega_{\mathrm{prel}}(x,t)$
and the transverse electric field $E_{z}(x,t)$ in the skin layer.
The initial undisturbed plasma frequency $\omega_{p0}$ is marked as the black dashed line.
Due to the relativistic effect, the effective plasma frequency is
largely decreased to be around a few THz, which is in the THz frequency gap range. The net transverse current modulated by this plasma frequency leads to the emission at the low THz frequencies. 

\begin{figure}
\includegraphics[width=0.45\textwidth
]{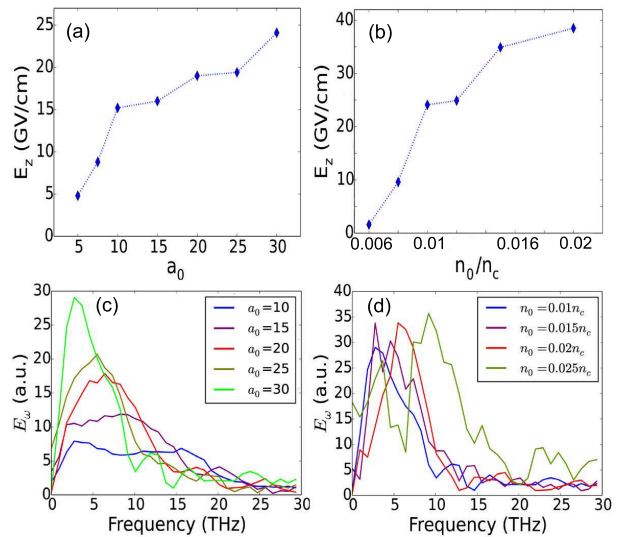}
\caption{\label{param} (Color online) (a)-(b): Peak THz field strength as a function of (a) the driven laser amplitude $a_0$ and (b) the initial plasma density $n_0$. (c)-(d): THz spectrum dependence on (c) $a_0$ and (d) $n_0$. The plasma density is $n_0=0.01n_c$ for panels (a) and (c). The laser amplitude is $a_0=30$ for panels (b) and (d).}
\end{figure}

We find the present scheme works for a wide range of laser and plasma
parameters. Figure~\ref{param} presents the results of a series of
simulations, where we take $n_{0}=0.01n_{c}$ in panels (a) and (c), and
$a_{0}=30$ in panels (b) and (d). Figs.~\ref{param} (a) and (b) shows the peak amplitude of the THz emission increases
with increase of the driver strength $a_{0}$ and the plasma density
$n_{0}$. Higher $a_{0}$ and $n_{0}$
both lead to a higher density spike in front of the laser pulse, which
results in a more efficient local depletion of the laser energy, and
thus a stronger radiation field. A rough scaling of the THz strength
dependence on the laser plasma parameters can be obtained using the
above 1D model. Assuming the source is sharply localized and exponentially
decays in the skin depth $l_{s}$, the integral of Eq.~(\ref{ezint})
can be approximated by $e_{z}^{T}\simeq\omega_{p}s_{z}/2\omega_{0}$.\cite{Lichters1996,Wu2008}
Concerning the source term, we further assume that the driver laser
is nonevolving. With the assumption of the quasistatic approximation,\cite{Sprangle1990}
the plasma fluid quantities can then be expressed as $n_{e}/n_{0}=[\gamma_{\perp}^{2}+(1+\phi^{2})]/2(1+\phi)^{2}$
and $\gamma=[\gamma_{\perp}^{2}+(1+\phi^{2})]/2(1+\phi)$, where $\gamma_{\perp}^{2}=1+a^{2}$
and $\phi$ is the electrostatic wake potential.\cite{Esarey2009}
In the near wavebreaking nonlinear limit $(1+\phi)\to1/\gamma_{p}$,
where the plasma Lorentz factor $\gamma_{p}=(1-v_{p}^{2}/c^{2})^{-1/2}$
with $v_{p}$ the phase velocity of the plasma wave. Then the THz
field can be obtained as $e_{z}^{T}\propto\omega_{p0}a_{z}^{R}\gamma_{p}$. 

The THz spectrum dependence on the laser intensity and plasma density are shown in Figs.~\ref{param}(c) and (d), respectively. With increasing the laser intensity, both the spectrum width and the central frequency decrease, and the peak spectral strength increases. The red shift of the central frequency is due to a lower relativistic plasma frequency, as a consequence of a larger $\gamma$-factor. Since the emitted pulse is single cycle, a lower-frequency (longer-wavelength) pulse has a longer temporal duration, and thus a narrower bandwidth. From Fig.~\ref{param}(d) one can see that the central THz frequency increases with increasing the plasma density, which is due to an increased plasma frequency. These results may offer a way to obtain a tunable THz source.  

\begin{figure}
\includegraphics[width=0.46\textwidth
]{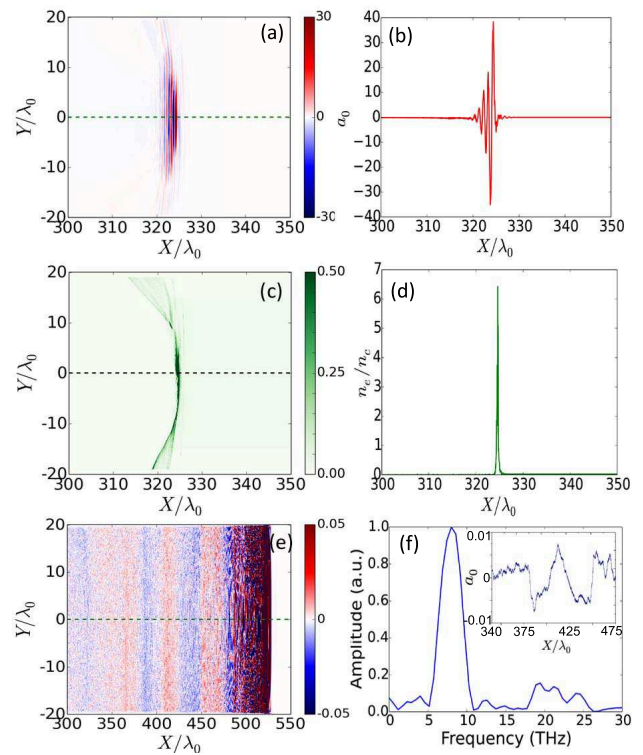}
\caption{\label{2d} (Color online) Results from 2D simulations. (a) A snapshot of the transverse electric field $eE_z/m_e \omega_0 c$ distribution and (b) the corresponding line-out plot along $Y=0$ at time $t=300T_0$. (c) A snapshot of the density $n_e$ distribution and (d) its line-out profile along $Y=0$ at $t=300T_0$. (e) A snapshot of the forward emission field $eE_z/m_e \omega_0 c$ distribution and (f) the Fourier spectrum of the field along $Y=0$ at $t=500T_0$. The inset of panel (f) shows the line-out plot of the THz field along $Y=0$, which was smoothed to get rid of the numerical noise.}
\end{figure}

In the following we present 2D PIC simulation results to check the
present scheme working in multi-dimensional geometries. The size of the simulation
box is $550\lambda_{0}\times60\lambda_{0}$ in the $xy$ plane and the grid step is $\lambda_{0}/50 \times \lambda_{0}/20$. The laser amplitude is
$a_{0}=30$ and the spot size is 15 $\mu$m FWHM. The plasma density
is $n_{0}=0.02n_{c}$. Figures \ref{2d}(a) and (c) show the 2D snapshots
of distribution of the transverse electric field $E_{z}$ and the
electron density $n_{e}$ at time $t=300T_{0}$, respectively. The
corresponding line-out plots along $Y=0$ are shown in Fig. \ref{2d}(b)
and (b). As in the 1D case, the laser pulse drives an electron density
spike at the front. The leading edge is eroding and becoming ultra-steep.
Figures \ref{2d}(e) and (f) show a snapshot of the electric field $E_{z}$,
i.e., the distribution of the forward emission, and its corresponding Fourier
spectrum along $Y=0$ at time $t=500T_{0}$. The inset of Fig.~\ref{2d}(f) shows the line-out plot of the THz electric field along $Y=0$. One can see that the
emission fields also have strong peaks in the THz frequency gap range
of $<$ 10 THz. The peak field strength is about 0.3 GV/cm. The reduction of the field strength can be attributed to multi-dimensional effects such as transverse electron spreading and
wave diffraction. In the 1D cases, the dense electron spike in front of the laser pulse maintains for a long time, provided the laser ponderomotive force is much larger than the electrostatic restoring force. In the multi-dimensional cases, the transverse spreading effects reduce both the average density and lifetime of the electron spike, which consequently impair the efficiency of the laser front etching and THz generation. Further increasing the plasma density and focusing
the THz emission may result in a higher field strength.

\section{Conclusions}
In conclusion, we have proposed a method to generate high field THz
radiation using a relativistic laser interaction with gaseous plasma targets.
An initially Gaussian laser pulse snow plows electrons forward to
form a sharp density spike in front of the pulse. Due to the local
pump depletion effect, the laser front is gradually etched to become
a step-function state. Plasma electrons gain a net residual momentum
after interaction with such a pulse and excite transverse oscillations
modulated by the relativistic plasma frequency, which emits the low
frequency THz pulses. This result opens the way for experimental approaches. 
This tunable THz source with a peak field strength reaching GV/cm level and spectrum mainly within the THz gap range, should be of interest for high-field THz science and applications.

\section*{Acknowledgments}
Z.-Y. C. acknowledges financial support from the China Scholarship Council (CSC). This work was supported by the Deutsche Forschungsgemeinschaft SFB TR 18 and by EU FP7 project EUCARD-2.  


%
%

%



\end{CJK}
\end{document}